\documentclass[%
 aip,
 jcp,%
 amsmath,amssymb,
preprint,%
]{revtex4-1}
\usepackage{natbib}
\usepackage{graphicx}
\usepackage{dcolumn}
\usepackage{bm}
\usepackage[hidelinks, urlcolor = blue]{hyperref}
\usepackage{url}
\usepackage{xcolor}

\usepackage{eso-pic}
\usepackage{fancyhdr}
\usepackage[headsep=2cm]{geometry}
\AddToShipoutPicture{%
  \AtTextUpperLeft{%
    \makebox(420,45)[lt]{%
      \footnotesize%
      \textbf{This manuscript was accepted by J. Chem. Phys. Click \href{https://aip.scitation.org/doi/10.1063/1.5086746}{here} to see the version of record.}\linktoVOR%
}}}
\newcommand{\linktoVOR}{}

\begin{document}


\title[A simple local expression for the prefactor in transition state theory]{A simple local expression for the prefactor in transition state theory}

\author{S. Kadkhodaei}
 \altaffiliation[Also at ]{Civil and Materials Engineering, University of Illinois, Chicago.}
 \email{sara\textunderscore kadkhodaei@brown.edu.}

\author{A. van de Walle}%
\affiliation{ 
School of Engineering, Brown University, Providence, Rhode Island 02912, USA 
}%

\date{\today}

\begin{abstract}
We present a simple and accurate computational technique to determine the frequency prefactor in harmonic transition state theory without necessitating full phonon density of states (DOS) calculations. The atoms in the system are partitioned into an ``active region'', where the kinetic process takes place, and an
``environment'' surrounding the active region. It is shown that the prefactor can be obtained by a partial phonon DOS calculation of the active region with a simple correction term accounting for the environment, under reasonable assumptions regarding atomic interactions. Convergence with respect to the size of the active region is investigated for different systems, as well as the reduction in computational costs when compared to full phonon DOS calculation. Additionally, we provide an open source implementation of the algorithm that can be added as an extension to LAMMPS software. 
\end{abstract}


\maketitle

\section{Introduction}
Kinetic processes play a major role in structure formation, microstructure development and materials properties, and diffusion is the underlying phenomenon in many of kinetic processes. 
%
The significant progress of atomic level simulations based on first-principles approaches and empirical interatomic potentials paved the way towards determining diffusivities by the use of computer simulations. Additionally, atomic level simulations provide fundamental insight into the atomic mechanisms in diffusion processes, and without relying on fitting to experimental data, they have predictive power for tuning of diffusion rate properties. 

A number of previous studies have used electronic structure calculations, model potential or a combination of both to obtain the diffusion coefficient from atomic level simulations. In general, interstitial or point defect jump frequency, which is the dominating mechanics for diffusion in most elemental crystals, is expressed by the equation
\begin{equation}\label{arrhenius}
\Gamma = {\nu}^*\exp\left(\frac{-\Delta E}{k_BT}\right),
\end{equation}
where $k_B$ is Boltzmann's constant, $T$ is temperature, $\Delta E$ represents the potential energy change needed to carry a defect from an initial equilibrium position to a transition state, and ${\nu}^*$ is the effective frequency associated with the vibration of the defect along the transition path. 
The atomic migration energy barrier calculation is more straightforward and can be obtained by using static total energy calculations. On the other hand, calculating the jump frequency needs more effort and depends on the change in the entropy associated with lattice vibration ($\Delta S$), which is due to the constriction or expansion of the diffusing atom's path during diffusion. 
%
One approach is to evaluate the diffusion rate directly from an analysis of the mean squared displacement of the diffusing atom during a molecular dynamics simulation~\cite{Tang1997,Mendelev2009}. While this method can deliver the pre-exponential factor in Eq.~\ref{arrhenius}, it is only practical at temperatures close to the melting point. Otherwise the diffusion events are too rare to be observed in the molecular dynamics simulation. 
 Milman et al. combined \textit{ab initio} molecular dynamics and thermodynamic integration to obtain the diffusion entropy change ($\Delta S$) for interstitial diffusion in silicon~\cite{Milman1993}. Other studies~\cite{Frank1996,Sandberg2002,Mantina2008,MANTINA2009} used the harmonic approximation within the transition state theory (TST) and expressed the effective jump frequency as the ratio of the product of normal vibration frequencies of the initial state of atomic migration to that of the non-imaginary normal frequencies of the transition state~\cite{VINEYARD1957},
\begin{equation}\label{effFreq}
 {\nu}^*= \frac{ \displaystyle \prod_{i=1}^{3N}\nu _{i}^{B}}{\displaystyle \prod_{i=2}^{3N}\nu _{i}^{A}},
\end{equation}
where $\nu^{B}_i$ and $\nu ^{A}_i$ are the normal vibrational frequencies (with $\nu_1^A $ denoting the omitted imaginary frequency associated with the unstable mode) in the initial basin and in the activated state, respectively, for a system of $N$ atoms.
(This expression assumes that the center of mass of the system is held by a fictitious
spring. This has the effect of assigning a fixed
stiffness to the rigid translation modes whose associated 
frequencies are then the same in the basin and in the activated states and thus
cancel out in the prefactor expression.)

The common approach to calculate the normal vibrational frequencies in Eq.~\ref{effFreq} has been the use of direct force-constant calculation~\cite{Wei1992}, where the computation cost scales cubically with the number of atoms in the system. This fact limited the previous atomic level simulations of lattice diffusion in solids based on first-principles calculations to system with at most a hundred of atoms~\cite{Frank1996,Sandberg2002,Mantina2008,MANTINA2009}, which make it difficult to study more complex processes, such as grain boundary diffusion. Relying only on empirical potential models in order to investigate diffusion processes in larger system (thousands of atoms) may not always be a viable alternative, due to complex multi-body bonding behavior. Deploying semi-empirical interatomic potential models by fitting to zero-temperature \textit{ab initio} data has been performed by Mendelev et.al~\cite{Mendelev2009}, but it requires the task of developing a potential model for various systems. Huang et al.~\cite{Huang2013} developed an ingenious linearly scalable approach by approximating the phonon DOS by the kernel polynomial method as an expansion in terms of Chebyshev polynomials to reduce the computational cost of the direct force-constant method.

Here, we use an approach that
facilitates transition rate calculations in large-scale systems
by replacing the computation and diagonalization of the Hessian of the whole system
by the formally equivalent calculation of a much smaller local effective Hessian.
For this purpose, we assume
that the entire system of atoms can be partitioned into an \textquotedblleft
active region\textquotedblright , where the kinetic process takes place, and
an \textquotedblleft environment\textquotedblright\ surrounding the active
region, whose elastic behavior is independent of the atomic geometry of the
active region. Our approach is perhaps most related, in its mathematical formalism, to the one of Ref. \onlinecite{Binder2015},
although their primary aim is to handle coarse-grained systems rather than fully atomistic simulations.
Therefore, our approach entails different methods for calculating the various input quantities needed for the rate calculations.

In this paper, we benchmark this local force-constant approach by applying it to a number of different kinetic processes using realistic energy models (ab initio hamiltonian, ReaxFF or embedded atom potentials): self-diffusion in bulk elemental phases of silver and aluminum, lithium interstitial diffusion in diamond cubic phase of silicon, and oxygen adsorbate surface diffusion on Pt(111) slab.

\section{Theory}
\label{maintheory}

\begin{figure}[h]
\centering
\includegraphics[width=\linewidth]{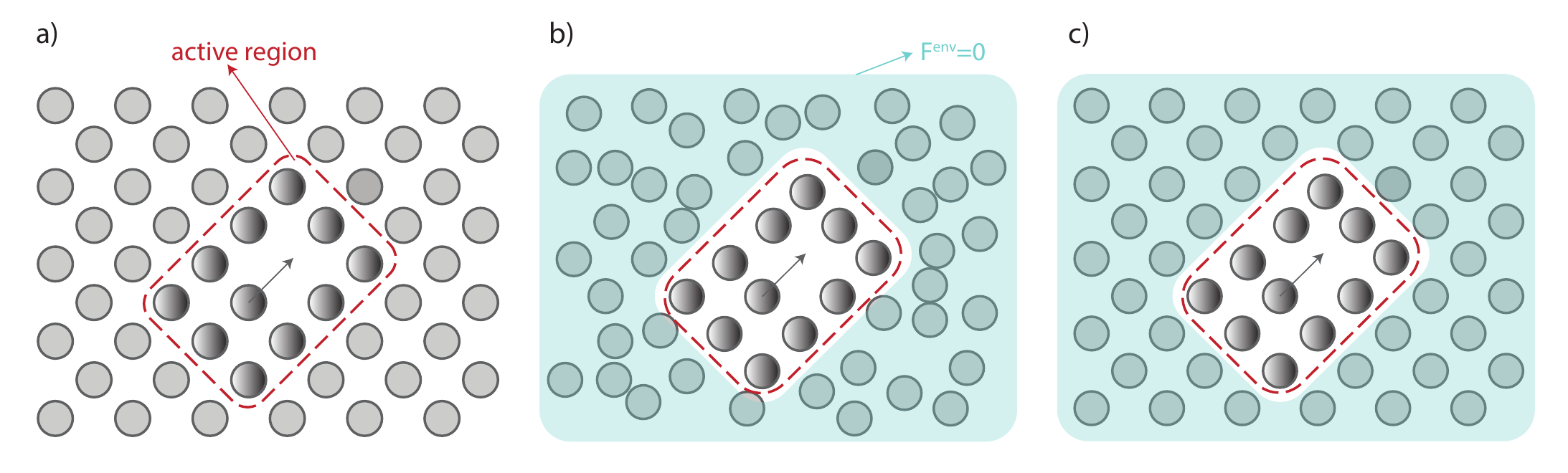}
\caption{a) Partitioning the atoms in a system into an ``active region'', where the kinetic process takes place, and an ``environment'' surrounding the active region. b) In the ``relaxed environment'' approach, the atoms in the environment are relaxed during force-constant calculation of active region atoms. c) In the ``fixed environment'' approach, no special treatment of environment atoms in conducted during the force-constant calculation of active region atoms.}
\label{fig:method}
\end{figure}

The partitioning of the system into an \textquotedblleft
active region\textquotedblright and an \textquotedblleft environment\textquotedblright\ (see Fig.~\ref{fig:method}) implies that
the system's Hessian (after suitable re-ordering of the atoms) can be partitioned in block form as well:
\begin{equation}
H^{A}=\left[ 
\begin{array}{cc}
A & C^{T } \\ 
C & E%
\end{array}%
\right] \text{ and }H^{B}=\left[ 
\begin{array}{cc}
B & D^{T } \\ 
D & E%
\end{array}%
\right] \label{partition1},
\end{equation}%
where $H^{A}$ and $H^{B}$ denote the Hessian in the activated state and in a
stable basin, respectively. The block $A$ and $B$ denote the respective Hessians of the active region
while block $E$ is the Hessian of the atoms in the environment, assumed to be independent of the configuration in the active
region. The $C$ and $D$ blocks represent the coupling between these two regions and can differ in the basin and activated state.

As noted in earlier work,\cite{Binder2015} a standard block-matrix determinant
identity implies that:%
\begin{eqnarray*}
\det H^{A} &=&\det \left( A-C^{T }E^{-1}C\right) \det \left( E\right) 
\\
\det H^{B} &=&\det \left( B-D^{T }E^{-1}D\right) \det \left( E\right) 
\end{eqnarray*}%
(a result shown in Appendix~\ref{sec:theory} for completeness). After introducing the
mass matrix $M$, we can then write the prefactor as:%
\begin{eqnarray*}
\nu ^{\ast } &=&\nu _{1}^{A}\frac{\prod_{i=1}^{3N}\nu _{i}^{B}}{%
\prod_{i=1}^{3N}\nu _{i}^{A}}=\nu _{1}^{A}\frac{\left( \det \left(
M^{-1/2}H^{B}M^{-1/2}\right) \right) ^{1/2}}{\left( \det \left(
M^{-1/2}H^{A}M^{-1/2}\right) \right) ^{1/2}} \\
&=&\left( \left( \nu _{1}^{A}\right) ^{2}\frac{\det \left( H^{B}\right) }{%
\det \left( H^{A}\right) }\right) ^{1/2}=\left( \left( \nu _{1}^{A}\right)
^{2}\frac{\det \left( B-D^{T }E^{-1}D\right) }{\det \left( A-C^{T
}E^{-1}C\right) }\right) ^{1/2}.
\end{eqnarray*}%
where $\nu _{1}^{A}$ is the imaginary frequency associated with the unstable
mode in the activated state. Terms of the form $B-D^{T }E^{-1}D$ and $%
A-C^{T }E^{-1}C$ can be interpreted as the local effective Hessian of
the active region, calculated while allowing the atoms in the environment to
relax until they experience no force (rather than being held fixed). These steps closely follow
earlier work,\cite{Binder2015} except for the specific way the terms $\left( \nu
_{1}^{A}\right) ^{2}$, $D^{T }E^{-1}D$ and $C^{T }E^{-1}C$ will be
computed in what follows.

In Ref. \onlinecite{Binder2015}, the authors seek to obtain a coarse-grained description of the system
in which the environment is removed but accounted for via a correction term $%
-D^{T }E^{-1}D$, which is computed once and then kept fixed throughout
the simulation (thus requiring $C=D$). They also aim to approximate $\left(
\nu _{1}^{A}\right) ^{2}$ only using the coarse-grained representation of
the system. In contrast, we do not seek to entirely remove the atoms of the
environment, which offers two advantages. First, it allows us circumvent the
requirement that $\left( \nu _{1}^{A}\right) ^{2}$ be well-approximated in a
coarse-grained system. Instead, we simply obtain $\left( \nu _{1}^{A}\right) ^{2}$
from%
\[
\left( \nu _{1}^{A}\right) ^{2} =\min_{\left\Vert u\right\Vert =1}u^{T }M^{-1/2}\tilde{H}%
^{A}M^{-1/2}u=\min_{\left\Vert M^{1/2}v\right\Vert =1}v^{T }\tilde{H}%
^{A}v,
\]%
a quantity that is closely related to what is already needed to find the
saddle point via, say, the Nudged Elastic Band \cite{neb1,neb2,neb3} or the
dimer \cite{dimer}\ method, again without requiring the calculation of the
full Hessian.

A second advantage is to allow us to relax the constraint $C=D$.
Relaxing this constraint does not take significantly more computational time
because one can use the relaxed environment geometry obtained in the
calculation of, say, $H^{B}$ as the starting point for the calculation of $%
H^{A}$. In a system where $C=D$ is a good approximation, the second
calculation will converge much faster.

The approach of Ref. \onlinecite{Binder2015}
is most useful when the same partitioning between the active
region and the environment can be used for all the transitions of interest while
ours is most useful when the transition events can occur anywhere in the
system so that the partitioning 
must be transition-dependent.

In the next sections, we show that the prefactor rates obtained by the above approach agree well with direct force-constant (full Hessian) calculation results, even for small sizes of the active region. Therefore, the use of the aforementioned approach significantly reduces the computation cost by limiting the number of degrees of freedom in force-constant calculations, while delivering accurate results comparable to direct force-constant calculation results. The additional cost of relaxing the atoms in the environment is much less that the burden of full Hessian calculations thanks to the high efficiency of modern conjugate gradient methods in high dimensions. Additionally, the cubically scaling cost of diagonalization of force-constant, which is the computational bottleneck in interatomic potential models, is significantly reduced by narrowing force-constant calculations efforts to atoms in the small active region.

\section{Computational Details}{\label{sec:details}} 
To test the applicability of the local force-constant method, we employ first-principles calculations in the framework of density functional theory (DFT) for vacancy hopping in bulk aluminum, for which there exist previous \textit{ab initio} calculations~\cite{Mantina2008}. However, the finite size effects cannot be excluded in the system sizes that are computationally accessible via DFT, i.e. at most tens of atoms. Therefore, we test the applicability of the model to systems with thousands of atoms by employing interatomic potential models for vacancy hopping in bulk silver, lithium diffusion in diamond cubic silicon, and oxygen surface diffusion on Pt(111) slab. To this end, we have implemented a C++ module in the Large-scale Atomic/Molecular Massively Parallel Simulator (LAMMPS) framework to add the functionality of local force-constant calculations for any interatomic potential. The source code of this module is open access and is available on the web~\cite{hesscode}. To invoke this new functionality, one has to simply put our C++ module code in the LAMMPS src directory and re-build it according to the LAMMPS website instructions~\cite{lammps-rebuild}.

For our first-principles calculations, we used the projector augmented wave (PAW) method~\cite{paw,paw_vasp}, as implemented in highly efficient Vienna Ab-initio Simulation Package (VASP)~\cite{vasp1,vasp2,vasp3,vasp4}, with energy cutoff of 300 eV. The generalized gradient approximation (GGA)~\cite{PERDEW1992} is used for exchange-correlation functional. The Brillouin zone integration is performed using a Monkhorst-Pack \textit{k}-mesh of $11\times 11\times 11$ for a 31-supercell of fcc aluminum. Tests indicate that this choice of first-principles calculation parameters ensures an energy accuracy of 0.1 meV/atom. 

For our interatomic potential calculations, we use an embedded atom (EAM) potential taken from Ref.~\cite{eamPot} to describe atomic interactions for silver and a reactive force field (ReaxFF) with parameters taken from Ref.~\cite{reaxffLiSi} to describe interatomic interactions of lithium and silicon. For the ReaxFF potential, we use the parameters derived in Ref.~\cite{reaxffLiSi}, where they are optimized against a DFT training set consisting of bulk c-$\text{Li}_{\text{x}}$Si and a-$\text{Li}_{\text{x}}$Si phases. A ReaxFF potential with optimized parameters to describe Pt-O system, taken from Ref.~\cite{reaxffPtO}, is used for oxygen diffusion on Pt(111) surface.

The number of atoms in our examples are limited to a couple of thousands atoms for the purpose of comparing our results against the direct force-constant calculation results, while the usage of our method is most advantageous where the number of atoms are in the order of millions and direct force-constant calculations are infeasible.

\section{Results}{\label{sec:results}} 
\subsection{Vacancy diffusion}
\begin{figure}[h]
\centering
\includegraphics[width=\linewidth]{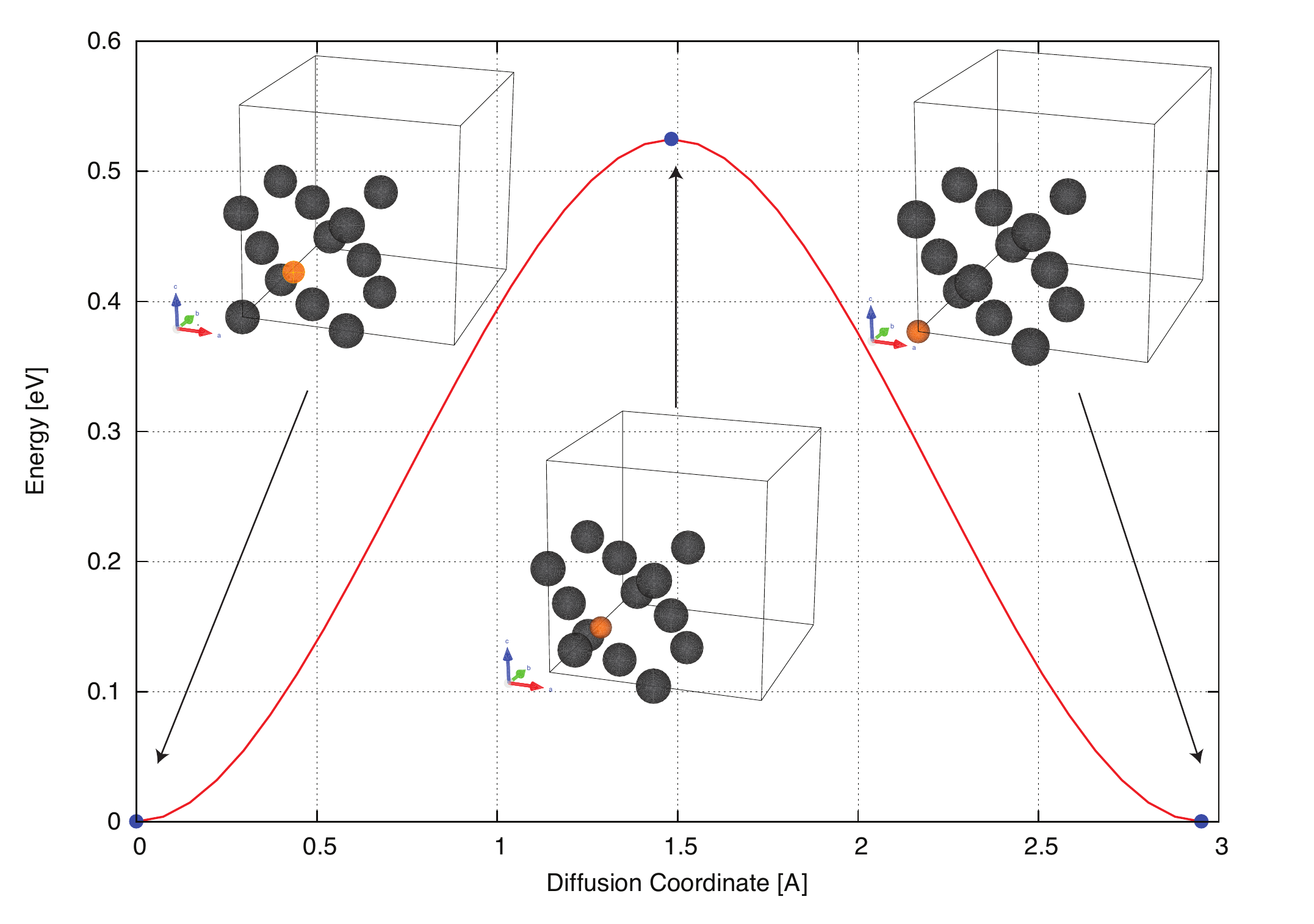}
\caption{Vacancy diffusion energy profile (DFT calculation) and the initial and saddle point configurations obtained using the nudged elastic band method. The vacancy is represented by a small sphere with different color than other atoms in the unit cell of fcc Al. The vacancy migrates from its initial position at [$\frac{a}{2},0,\frac{a}{2}$] along the saddle point configuration to the final position on [0 0 0], where a is the lattice constant and equals 4.09 $\AA$ in our calculations. In other words, the atom positioned at [0 0 0] in the initial composition migrates to[$\frac{a}{2},0,\frac{a}{2}$] position, passing through the [$\frac{a}{4},0,\frac{a}{4}$] position. Only atoms in fcc unit cell are represented for the purpose of clarity.}
\label{fig:vacancyMigration}
\end{figure}

\begin{figure}[h]
\centering
\includegraphics[width=\linewidth]{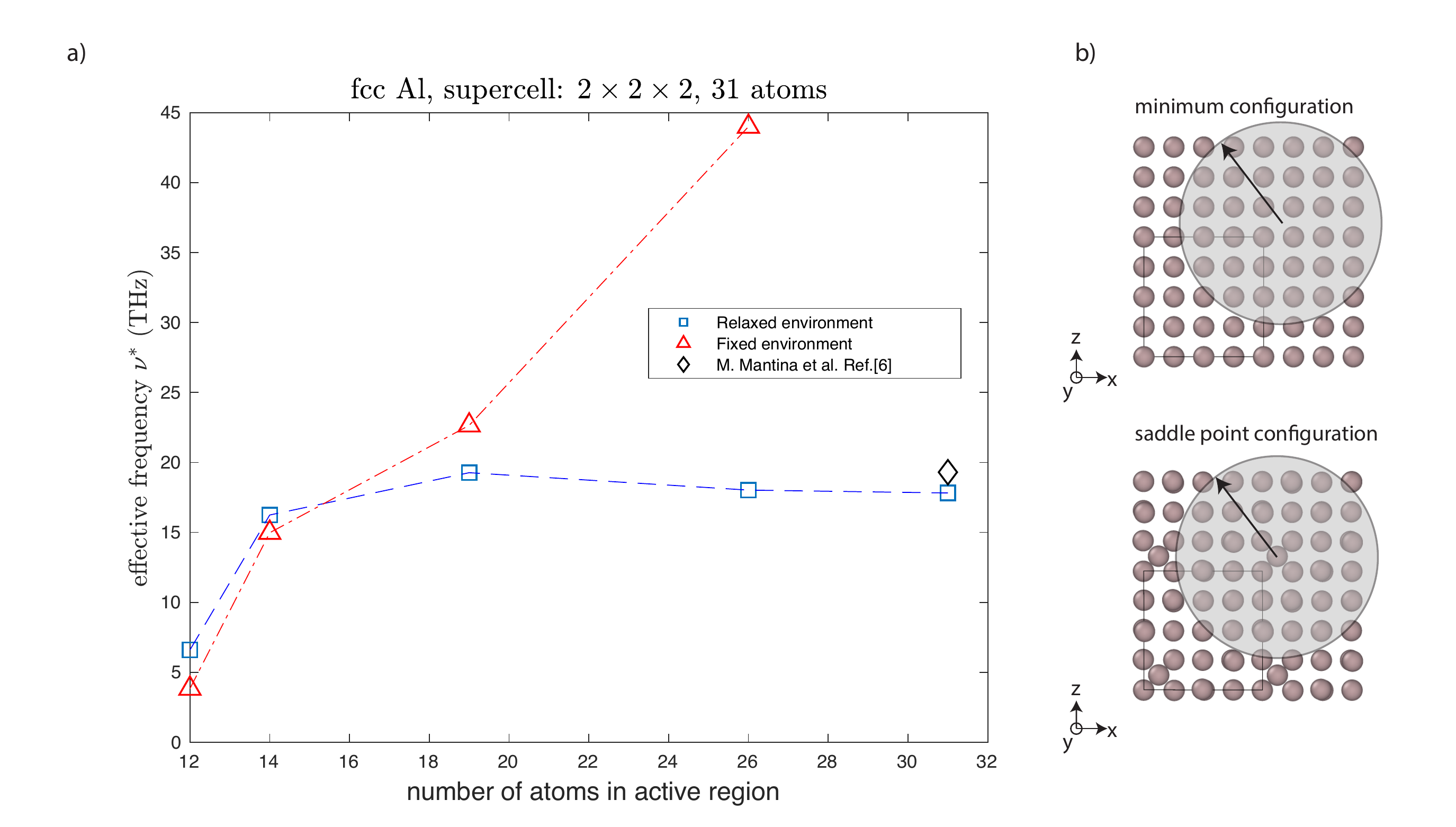}
\caption{a) Effective jump frequency of vacancy hopping in aluminum obtained by a partial Hessian calculation using DFT. The x-axis indicates the number of atoms included in the active region, which is a sphere originated at the position of migrating atom in the saddle point configuration. The blue squares are the prefactor values obtained by our method, where the atoms in the environment are relaxed during the force constant calculation. The red triangles are the prefactor rates obtained by a crude local force-constant calculation, where the atoms in the environment are fixed. The prefactor value reported in Ref.~\cite{Mantina2008} using the full Hessian calculation is represented by a black diamond. b) Minimum and saddle point configurations in a $2\times2\times2$ supercell of fcc Al. The active region is indicated as a sphere that encompasses atoms that are included in the force-constant calculations.}
\label{fig:dft_fcc_al}
\end{figure}

We first demonstrate the local force-constant method by computing the effective jump frequency (pre-exponential factor), $\nu^*$, for vacancy hopping in bulk aluminum using DFT calculations. The minimum and saddle point configurations are obtained using the nudged elastic band method as implemented in VASP~\cite{neb1,neb2}. The energy barrier we obtained is 0.52, and when a correction value of 0.05 for the ``internal surface'' effect~\cite{surfaceEffect} is added to it, the final value of 0.57 eV is in  agreement with the vacancy migration energy of 0.57 eV reported in Ref.\cite{Mantina2008}. Fig.~\ref{fig:vacancyMigration} represents the vacancy migration and its energy profile in fcc aluminum. The atomic configurations are visualized using the VESTA software~\cite{VESTA}. Having the minimum and saddle point configurations, we calculated the prefactor for different number of atoms in the active region employing the proposed method, as illustrated in Fig.~\ref{fig:dft_fcc_al}. Active region is defined as a sphere originated at the position of migrating atom in the saddle point configuration, as visualized by OVITO software~\cite{ovito} and represented in Fig.~\ref{fig:dft_fcc_al}. A linear increase in the radius of the active region results in a cubic increase in the number of atoms inside the active region. As shown in this figure, our results are compared  with a crude local approximation of force-constant, where the effect of atoms in the environment is not accounted for. In other words, in this crude approximation the force-constant calculation is conducted for atoms in the active region while other atoms in the system are fixed. We call this approach the ``fixed environment'' method, as compared to the proposed method of ``relaxed environment''. The results of the ``relaxed environment'' method indicate a better convergence compared to the ``fixed environment'' local force-constant calculation. The calculated prefactor values are in perfect agreement with the value reported in Ref.~\cite{Mantina2008} that is obtained using the direct force-constant calculation. 

\begin{figure}[h]
\centering
\includegraphics[width=\linewidth]{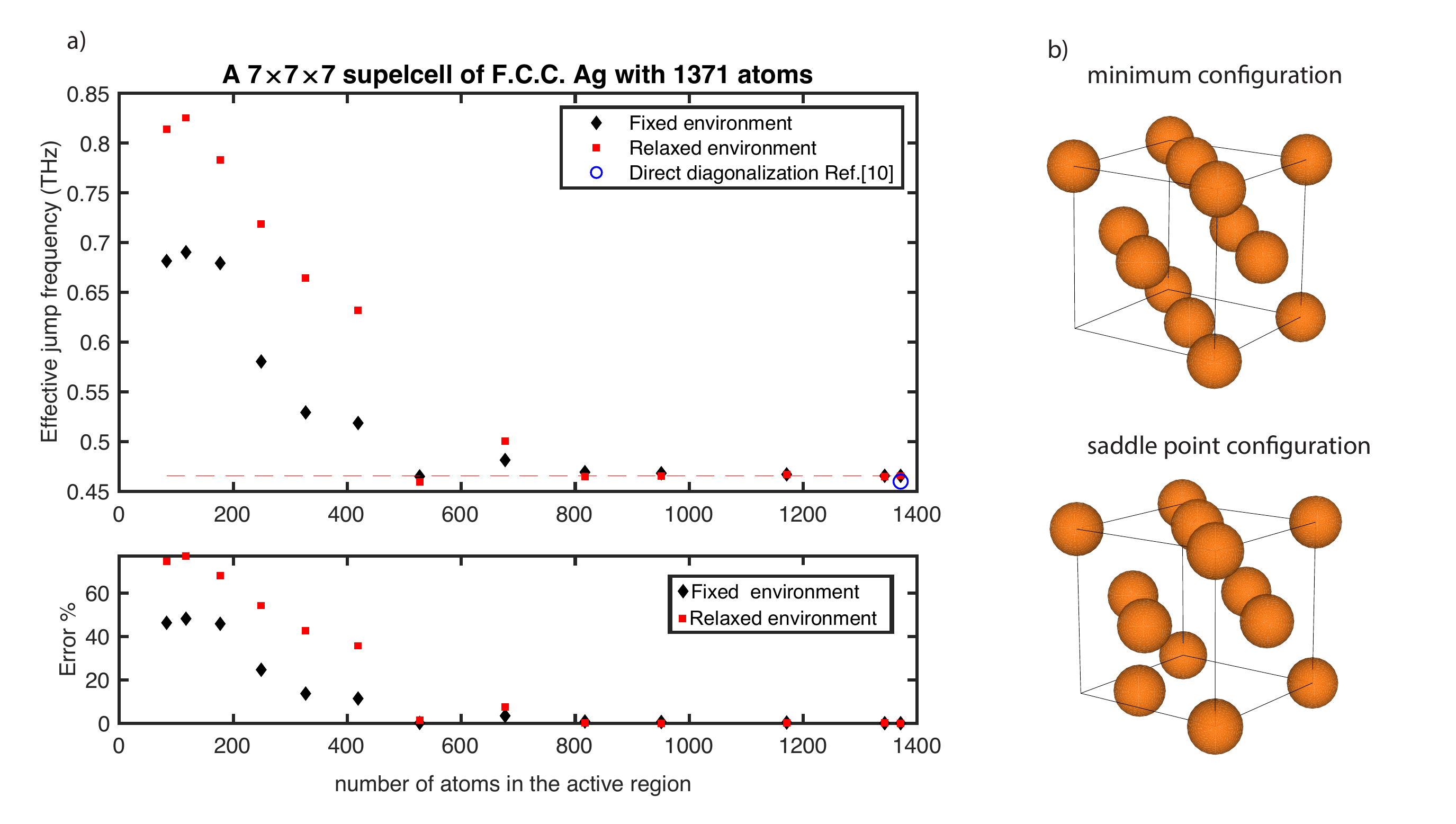}
\caption{a) Effective jump frequency of vacancy hopping in silver obtained by a our method using EAM potential. The x-axis indicates the number of atoms included in the active region, which is a sphere originated at the position of migrating atom in the saddle point configuration. The red squares are the prefactor values obtained by our method, where the atoms in the environment are relaxed during the force constant calculation. The black diamonds are the prefactor rates obtained by the ``fixed environment'' local force-constant calculation, where the atoms in the environment are fixed. The prefactor value reported in Ref.~\cite{Huang2013} is represented by a blue circle. b) Minimum and saddle point configurations for vacancy jump indicated in a unit cell of fcc Ag.}
\label{fig:eam_fcc_ag}
\end{figure}

Next, we demonstrate our method by computing the effective jump frequency for vacancy hopping in a larger system, a $7\times 7\times 7$ supercell of the conventional cubic cell of fcc silver with a vacancy consisting of 1371 atoms, using the EAM potential model. We used nudged elastic band method, as implemented in LAMMPS~\cite{neblammps1,neblammps2}, to obtain the minimum and transition state configurations. The initial and saddle point configurations for vacancy diffusion in Ag are visualized using VESTA software~\cite{VESTA} and are illustrated in Fig.~\ref{fig:eam_fcc_ag}. The energy barrier obtained from our calculations is 0.66 eV, in good agreement with the value of 0.65 eV reported in Ref.~\cite{Huang2013}. As presented in Fig.~\ref{fig:eam_fcc_ag}, we calculated the prefactor rate using our method of local force-constant calculation, where the atoms in the environment are relaxed during force-constant calculation of active region atoms. 
We compared our method with the case of fixing atoms in the environment during force-constant calculations. This comparison indicates that the ``fixed environment'' approach results happen, in this case, to converge faster due to fortuitous cancellation of errors. As discussed in Supplementary Information, while the partial Hessian calculated in the ``fixed environment'' approach ($A$ and $B$ in Eq.~\ref{partition1}) does not include the effect of atoms in environment, the ratio of partial Hessian determinants ($\frac{\det(B)}{\det(A)}$) happens to be in a good agreement with effective Hessian determinants ratio ($\frac{\det(\tilde{H}^{B})}{\det(\tilde{H}^{A})}$) (see Supplementary Information for a detailed discussion). This case is rather exceptional, however, and the remaining examples we consider below indicate that ``relaxed environment'' approach more typically performs better.
As shown in Fig.~\ref{fig:eam_fcc_ag}, even for small sizes of active region, the error magnitude is still fairly small (less than a factor 2), which is typically smaller than errors arising from other sources (such the use of empirical potential models). This fact enables the use of our method for small sizes of active region, which significantly reduces the computational costs involved in vibration frequency calculations. 

\subsection{Interstitial diffusion}
\begin{figure}[h]
\centering
\includegraphics[width=\linewidth]{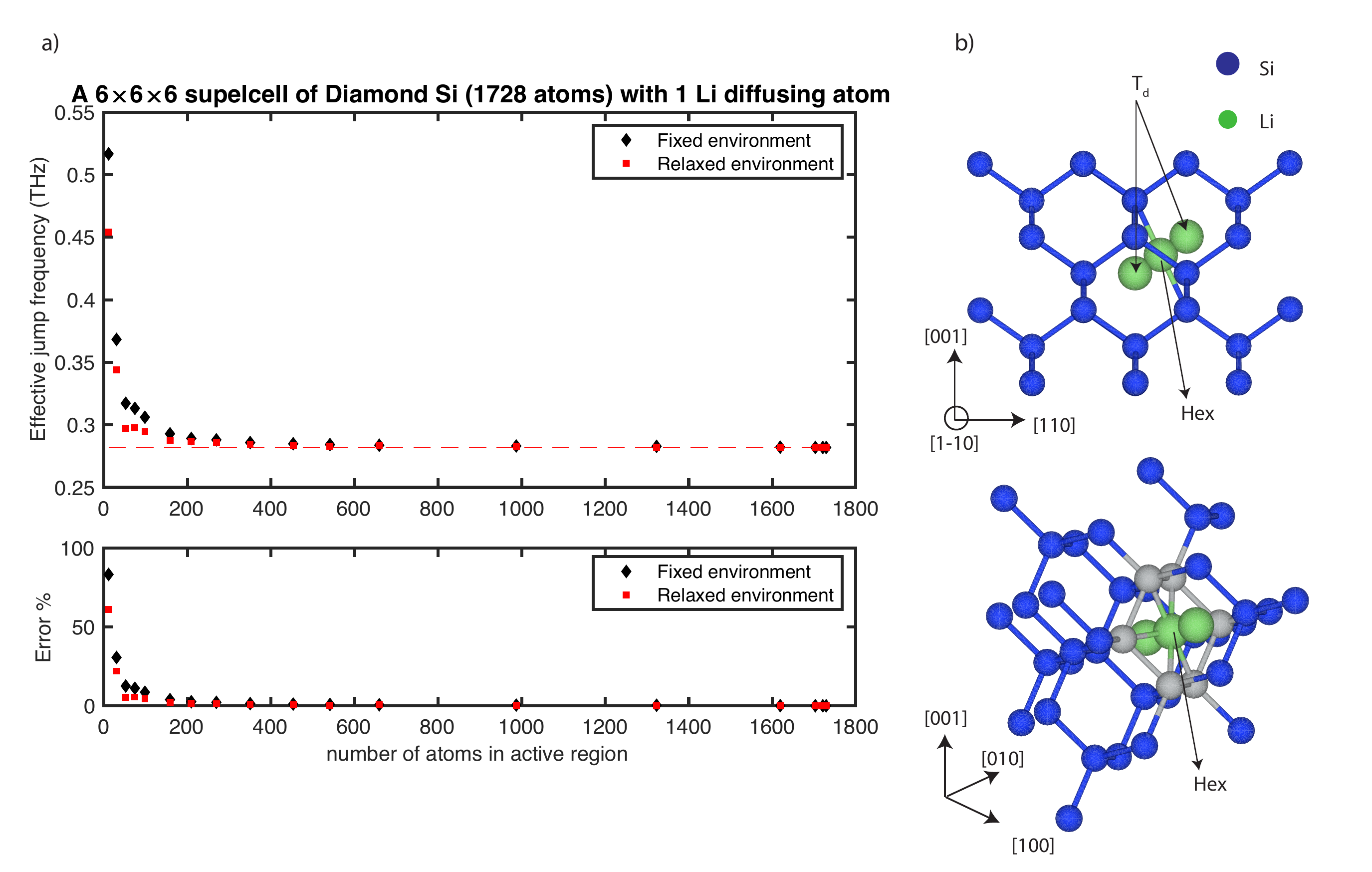}
\caption{a) Effective jump frequency of lithium interstitial diffusion in cubic diamond silicon, obtained by a our method using ReaxFF potential. The x-axis indicates the number of atoms included in the active region, which is a sphere originated at the mean position of Li atom in the minimum and saddle point configurations. The red squares are the prefactor values obtained by our method, where the atoms in the environment are relaxed during the force constant calculation. The black diamonds are the prefactor rates obtained by the ``fixed environment'' local force-constant calculation, where the atoms in the environment are fixed. b) Representation of minimum and saddle point configurations of Li diffusion in a 8-atom cubic diamond unitcell of Si. The tetrahedral ($T_d$) and  hexagonal (Hex) interstitial positions are depicted, corresponding to minimum and saddle point configurations, respectively. In the lower plot, the 6 nearest neighbor atoms for Hex interstitial are represented by a different color for the purpose of clarity.}
\label{fig:reax_LiSi}
\end{figure}

Here, we demonstrate the effective jump frequency for diffusion of a single Li atom through a diamond cubic supercell of 1728 Si atoms, corresponding to a $6\times6\times6$ supercell of 8-atom unitcell, using the ReaxFF potential model. The thermodynamically favorable site for Li atom insertion in cubic diamond Si is a tetrahedral site ($T_d$) at $b_0$ distance away from a Si atom in the opposite direction of Si nearest neighbor, where $b_0$ is the bond length in a cubic Si structure. The diffusion process consists of a Li atom moving between two adjacent interstitial $T_d$ sites passing through the hexagonal (Hex) interstitial position. Our calculation results in a lattice constant of 5.46 $\AA$ for c-Si (and a corresponding $b_0$ value of 2.37 $\AA$), in good agreement with values reported in Ref.~\cite{diffDFTSi, diffusionReaxSi}, using both DFT and ReaxFF calculations. We calculated the energy of Li atom in $T_d$  and Hex positions and the energy difference is 0.63 eV, in good agreement with previous energy barrier values reported in the literature~\cite{diffDFTSi, diffusionReaxSi}. Li interstitial at $T_d$ and Hex positions are indicated in Fig.~\ref{fig:reax_LiSi} using VESTA. The prefactor rates are calculated with our method and compared with the ``fixed environment'' approach and the direct force constant calculation results. As indicated in Fig.~\ref{fig:reax_LiSi}, the discrepancy between our results and full Hessian calculation results remains relevantly small, even for small number of atoms in the active region, confirming the validity of our approach and its effectiveness on reducing the computational cost. We also observe that the ``fixed environment'' results are very similar to the ``relaxed environment'' results, slightly underperforming relative to the ``relaxed environment'' approach.
\subsection{Surface diffusion}
\begin{figure}[h]
\centering
\includegraphics[width=\linewidth]{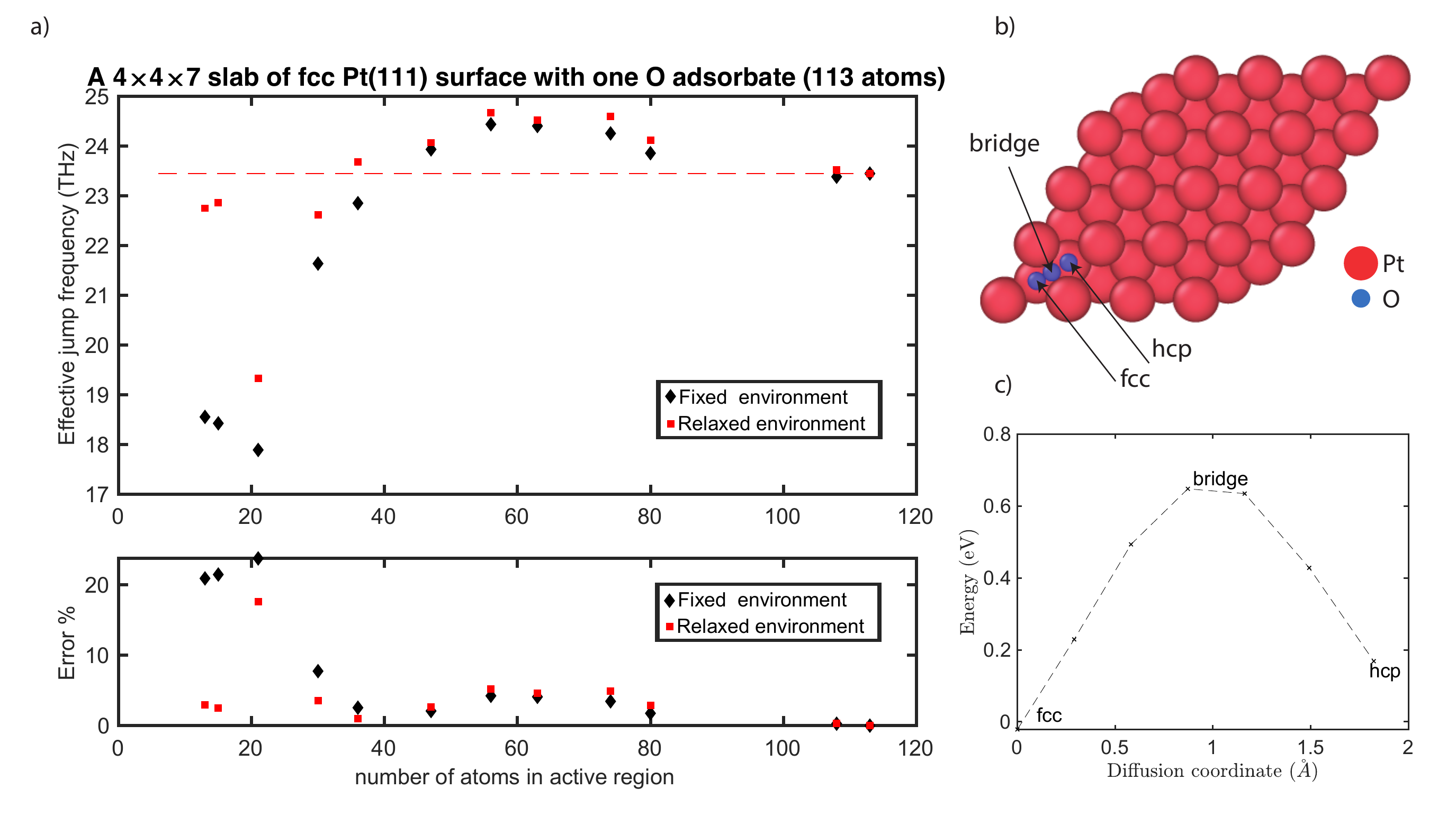}
\caption{a) Effective jump frequency of oxygen adsorbate diffusion on platinum (111) surface, obtained by a our method using ReaxFF potential. The x-axis indicates the number of atoms included in the active region, which is a sphere originated at the bridge site position. The red squares are the prefactor values obtained by our method, where the atoms in the environment are relaxed during the force constant calculation. The black diamonds are the prefactor rates obtained by the ``fixed environment'' local force-constant calculation, where the atoms in the environment are fixed. b) Representation of fcc, bridge and hcp adsorption site on Pt(111) surface. c) Energy profile for oxygen bridge-diffusion on Pt(111) surface. The x-axis is the diffusion coordinate distance in $\AA$ and the y-axis is the energy in eV.}
\label{fig:reax_PtO}
\end{figure}

We finally represent the effective jump frequency rates for oxygen adsorbate diffusion on platinum (111) surface. We use a 7-layer thick Pt(111) surface with each layer consisting of $4\times4$ atoms, including 112 platinum atoms in total. This simulation cell size is chosen since there exist previous studies for the diffusion profile for this system size using ReaxFF~\cite{reaxffPtO}. The high symmetry adsorption site on a fcc(111) surface include the fcc, hcp, top and bridge sits, as indicated in Fig.~\ref{fig:reax_PtO}. We calculate the prefactor rates for the bridge-diffusion, which consists of an adsorbate moving from an fcc site to the nearby hcp site, passing through the bridge site. To isolate the oxygen adsorbate from the adjacent simulation cells, we included a 30$\AA$ vacuum region in the simulation cell. We conducted nudged elastic band calculations in LAMMPS to obtain the bridge-diffusion path for oxygen surface diffusion on Pt(111). The diffusion energy profile along with the initial, transition and final states are represented in Fig.~\ref{fig:reax_PtO}, visualized by OVITO. The calculated forward (fcc-to-hcp) and backward (hcp-to-fcc) energy barriers are 0.65 eV and 0.48 eV, in good agreement with values reported in Ref.~\cite{reaxffPtO}. The prefactor rates are then calculated for different number of atom in the active region, which is a sphere originated at bridge site position. Our results indicate very good convergence to the full Hessian calculation results and the range of error remains as small as 20\%. The ``fixed environment`` results slightly underperform relative to our approach, although both methods give reliable results.. 

\section{Discussion}{\label{sec:discussion}}
Our results bring out two key messages. First, one can rigorously define the harmonic TST prefactor in terms of local effective Hessians, where the effect of the environment enters via a simple correction term associated with static elastic relaxations of the environment. Second, we find, through numerical simulations in a wide range of systems, that the corrections due to relaxing the environment tend to be similar in the stable and saddle point configurations, thus justifying an even simpler ``fixed environment" approach. This partial error cancellation (that occur beyond what our theoretical model predicts) is analyzed in more detail the Supplementary Information.

These observations open the way to simple, local approaches to computing the TST prefactor. Although we do not explore this possibility in detail here, our observations also suggest efficient hybrid schemes using three different regions: (i) an ``active" region whose Hessian is explicitly calculated; (ii) an intermediate region surrounding it where atoms are relaxed during the computation of the effective Hessian of the active region; and (iii) an environment where all atoms are fixed.

\section{Conclusion}
In this paper, we examined an approach based on a local force-constant calculation in the vicinity of where a kinetic process takes place to obtain the prefactor in transition state theory. While the standard force constant calculation procedure is performed on an ``active region'', the overall effect of ``environment'' atoms is accounted for by relaxing them during this procedure. We demonstrated that the effective jump frequencies obtain using the proposed method rapidly converges to the full force-constant calculation results for a number of diffusion processes, validating the theory employed herein. We also found empirically that a faster and easier local force-constant approach, where the overall effect of atoms in the ``environment'' is disregarded by fixing them during the local force-constant process, results in very similar prefactor rates due to the cancellation of errors. The proposed approach significantly reduce the computation efforts involved in force-constant calculation by restricting the number of atoms to the vicinity of the kinetic event. 

\section{Supplementary Information}
See supplementary information for a detailed discussion on the cancellation of errors in frequency prefactor calculation for the ``fixed environment'' method.

\section{Acknowledgments}
This work is supported by the Office of Naval Research under grant N00014-17-1-2202 and by Brown University through the use of the facilities of its Center
for Computation and Visualization. This work used the Extreme Science and Engineering Discovery Environment (XSEDE), which is supported by National Science Foundation grant number ACI-1548562.

\appendix
\section{Simple derivation of a block matrix determinant identity.}{\label{sec:theory}} 
First observe that
\[
\det \left( \left[ 
\begin{array}{cc}
I & 0 \\ 
X & L%
\end{array}%
\right] \right) =\det \left( L\right) ,
\]
where $I$ is the identity matrix, $0$ is a matrix of zeros, while $X$ and $L$ are conformable arbitrary matrices.
We can then write (in the notation of Section \ref{maintheory})%
\begin{eqnarray*}
\det \left( H_{A}\right)  &=&\det \left( \left[ 
\begin{array}{cc}
A & C^{T } \\ 
C & E%
\end{array}%
\right] \right) \det \left( E^{-1}\right) \det \left( E\right)  \\
&=&\det \left( \left[ 
\begin{array}{cc}
A & C^{T } \\ 
C & E%
\end{array}%
\right] \right) \det \left( \left[ 
\begin{array}{cc}
I & 0 \\ 
-E^{-1}C & E^{-1}%
\end{array}%
\right] \right) \det \left( E\right)  \\
&=&\det \left( \left[ 
\begin{array}{cc}
A & C^{T } \\ 
C & E%
\end{array}%
\right] \left[ 
\begin{array}{cc}
I & 0 \\ 
-E^{-1}C & E^{-1}%
\end{array}%
\right] \right) \det \left( E\right)  \\
&=&\det \left( \left[ 
\begin{array}{cc}
A-C^{T }E^{-1}C & C^{T }E^{-1} \\ 
0 & I%
\end{array}%
\right] \right) \det \left( E\right)  \\
&=&\det \left( A-C^{T }E^{-1}C\right) \det \left( E\right) .
\end{eqnarray*}

Similarly, $\det \left( H_{B}\right) =\det \left( B-D^{T
}E^{-1}D\right) \det \left( E\right) .$


\begin{thebibliography}{35}%
\makeatletter
\providecommand \@ifxundefined [1]{%
 \@ifx{#1\undefined}
}%
\providecommand \@ifnum [1]{%
 \ifnum #1\expandafter \@firstoftwo
 \else \expandafter \@secondoftwo
 \fi
}%
\providecommand \@ifx [1]{%
 \ifx #1\expandafter \@firstoftwo
 \else \expandafter \@secondoftwo
 \fi
}%
\providecommand \natexlab [1]{#1}%
\providecommand \enquote  [1]{``#1''}%
\providecommand \bibnamefont  [1]{#1}%
\providecommand \bibfnamefont [1]{#1}%
\providecommand \citenamefont [1]{#1}%
\providecommand \href@noop [0]{\@secondoftwo}%
\providecommand \href [0]{\begingroup \@sanitize@url \@href}%
\providecommand \@href[1]{\@@startlink{#1}\@@href}%
\providecommand \@@href[1]{\endgroup#1\@@endlink}%
\providecommand \@sanitize@url [0]{\catcode `\\12\catcode `\$12\catcode
  `\&12\catcode `\#12\catcode `\^12\catcode `\_12\catcode `\%12\relax}%
\providecommand \@@startlink[1]{}%
\providecommand \@@endlink[0]{}%
\providecommand \url  [0]{\begingroup\@sanitize@url \@url }%
\providecommand \@url [1]{\endgroup\@href {#1}{\urlprefix }}%
\providecommand \urlprefix  [0]{URL }%
\providecommand \Eprint [0]{\href }%
\providecommand \doibase [0]{http://dx.doi.org/}%
\providecommand \selectlanguage [0]{\@gobble}%
\providecommand \bibinfo  [0]{\@secondoftwo}%
\providecommand \bibfield  [0]{\@secondoftwo}%
\providecommand \translation [1]{[#1]}%
\providecommand \BibitemOpen [0]{}%
\providecommand \bibitemStop [0]{}%
\providecommand \bibitemNoStop [0]{.\EOS\space}%
\providecommand \EOS [0]{\spacefactor3000\relax}%
\providecommand \BibitemShut  [1]{\csname bibitem#1\endcsname}%
\let\auto@bib@innerbib\@empty
\bibitem [{\citenamefont {Tang}\ \emph {et~al.}(1997)\citenamefont {Tang},
  \citenamefont {Colombo}, \citenamefont {Zhu},\ and\ \citenamefont {Diaz de~la
  Rubia}}]{Tang1997}%
  \BibitemOpen
  \bibfield  {author} {\bibinfo {author} {\bibfnamefont {M.}~\bibnamefont
  {Tang}}, \bibinfo {author} {\bibfnamefont {L.}~\bibnamefont {Colombo}},
  \bibinfo {author} {\bibfnamefont {J.}~\bibnamefont {Zhu}}, \ and\ \bibinfo
  {author} {\bibfnamefont {T.}~\bibnamefont {Diaz de~la Rubia}},\ }\href
  {\doibase 10.1103/PhysRevB.55.14279} {\bibfield  {journal} {\bibinfo
  {journal} {Phys. Rev. B}\ }\textbf {\bibinfo {volume} {55}},\ \bibinfo
  {pages} {14279} (\bibinfo {year} {1997})}\BibitemShut {NoStop}%
\bibitem [{\citenamefont {Mendelev}\ and\ \citenamefont
  {Mishin}(2009)}]{Mendelev2009}%
  \BibitemOpen
  \bibfield  {author} {\bibinfo {author} {\bibfnamefont {M.~I.}\ \bibnamefont
  {Mendelev}}\ and\ \bibinfo {author} {\bibfnamefont {Y.}~\bibnamefont
  {Mishin}},\ }\href {\doibase 10.1103/PhysRevB.80.144111} {\bibfield
  {journal} {\bibinfo  {journal} {Phys. Rev. B}\ }\textbf {\bibinfo {volume}
  {80}},\ \bibinfo {pages} {144111} (\bibinfo {year} {2009})}\BibitemShut
  {NoStop}%
\bibitem [{\citenamefont {Milman}\ \emph {et~al.}(1993)\citenamefont {Milman},
  \citenamefont {Payne}, \citenamefont {Heine}, \citenamefont {Needs},
  \citenamefont {Lin},\ and\ \citenamefont {Lee}}]{Milman1993}%
  \BibitemOpen
  \bibfield  {author} {\bibinfo {author} {\bibfnamefont {V.}~\bibnamefont
  {Milman}}, \bibinfo {author} {\bibfnamefont {M.~C.}\ \bibnamefont {Payne}},
  \bibinfo {author} {\bibfnamefont {V.}~\bibnamefont {Heine}}, \bibinfo
  {author} {\bibfnamefont {R.~J.}\ \bibnamefont {Needs}}, \bibinfo {author}
  {\bibfnamefont {J.~S.}\ \bibnamefont {Lin}}, \ and\ \bibinfo {author}
  {\bibfnamefont {M.~H.}\ \bibnamefont {Lee}},\ }\href {\doibase
  10.1103/PhysRevLett.70.2928} {\bibfield  {journal} {\bibinfo  {journal}
  {Phys. Rev. Lett.}\ }\textbf {\bibinfo {volume} {70}},\ \bibinfo {pages}
  {2928} (\bibinfo {year} {1993})}\BibitemShut {NoStop}%
\bibitem [{\citenamefont {Frank}\ \emph {et~al.}(1996)\citenamefont {Frank},
  \citenamefont {Breier}, \citenamefont {Els\"asser},\ and\ \citenamefont
  {F\"ahnle}}]{Frank1996}%
  \BibitemOpen
  \bibfield  {author} {\bibinfo {author} {\bibfnamefont {W.}~\bibnamefont
  {Frank}}, \bibinfo {author} {\bibfnamefont {U.}~\bibnamefont {Breier}},
  \bibinfo {author} {\bibfnamefont {C.}~\bibnamefont {Els\"asser}}, \ and\
  \bibinfo {author} {\bibfnamefont {M.}~\bibnamefont {F\"ahnle}},\ }\href
  {\doibase 10.1103/PhysRevLett.77.518} {\bibfield  {journal} {\bibinfo
  {journal} {Phys. Rev. Lett.}\ }\textbf {\bibinfo {volume} {77}},\ \bibinfo
  {pages} {518} (\bibinfo {year} {1996})}\BibitemShut {NoStop}%
\bibitem [{\citenamefont {Sandberg}, \citenamefont {Magyari-K\"ope},\ and\
  \citenamefont {Mattsson}(2002)}]{Sandberg2002}%
  \BibitemOpen
  \bibfield  {author} {\bibinfo {author} {\bibfnamefont {N.}~\bibnamefont
  {Sandberg}}, \bibinfo {author} {\bibfnamefont {B.}~\bibnamefont
  {Magyari-K\"ope}}, \ and\ \bibinfo {author} {\bibfnamefont {T.~R.}\
  \bibnamefont {Mattsson}},\ }\href {\doibase 10.1103/PhysRevLett.89.065901}
  {\bibfield  {journal} {\bibinfo  {journal} {Phys. Rev. Lett.}\ }\textbf
  {\bibinfo {volume} {89}},\ \bibinfo {pages} {065901} (\bibinfo {year}
  {2002})}\BibitemShut {NoStop}%
\bibitem [{\citenamefont {Mantina}\ \emph {et~al.}(2008)\citenamefont
  {Mantina}, \citenamefont {Wang}, \citenamefont {Arroyave}, \citenamefont
  {Chen}, \citenamefont {Liu},\ and\ \citenamefont {Wolverton}}]{Mantina2008}%
  \BibitemOpen
  \bibfield  {author} {\bibinfo {author} {\bibfnamefont {M.}~\bibnamefont
  {Mantina}}, \bibinfo {author} {\bibfnamefont {Y.}~\bibnamefont {Wang}},
  \bibinfo {author} {\bibfnamefont {R.}~\bibnamefont {Arroyave}}, \bibinfo
  {author} {\bibfnamefont {L.~Q.}\ \bibnamefont {Chen}}, \bibinfo {author}
  {\bibfnamefont {Z.~K.}\ \bibnamefont {Liu}}, \ and\ \bibinfo {author}
  {\bibfnamefont {C.}~\bibnamefont {Wolverton}},\ }\href {\doibase
  10.1103/PhysRevLett.100.215901} {\bibfield  {journal} {\bibinfo  {journal}
  {Phys. Rev. Lett.}\ }\textbf {\bibinfo {volume} {100}},\ \bibinfo {pages}
  {215901} (\bibinfo {year} {2008})}\BibitemShut {NoStop}%
\bibitem [{\citenamefont {Mantina}\ \emph {et~al.}(2009)\citenamefont
  {Mantina}, \citenamefont {Wang}, \citenamefont {Chen}, \citenamefont {Liu},\
  and\ \citenamefont {Wolverton}}]{MANTINA2009}%
  \BibitemOpen
  \bibfield  {author} {\bibinfo {author} {\bibfnamefont {M.}~\bibnamefont
  {Mantina}}, \bibinfo {author} {\bibfnamefont {Y.}~\bibnamefont {Wang}},
  \bibinfo {author} {\bibfnamefont {L.}~\bibnamefont {Chen}}, \bibinfo {author}
  {\bibfnamefont {Z.}~\bibnamefont {Liu}}, \ and\ \bibinfo {author}
  {\bibfnamefont {C.}~\bibnamefont {Wolverton}},\ }\href {\doibase
  https://doi.org/10.1016/j.actamat.2009.05.006} {\bibfield  {journal}
  {\bibinfo  {journal} {Acta Materialia}\ }\textbf {\bibinfo {volume} {57}},\
  \bibinfo {pages} {4102 } (\bibinfo {year} {2009})}\BibitemShut {NoStop}%
\bibitem [{\citenamefont {Vineyard}(1957)}]{VINEYARD1957}%
  \BibitemOpen
  \bibfield  {author} {\bibinfo {author} {\bibfnamefont {G.~H.}\ \bibnamefont
  {Vineyard}},\ }\href {\doibase https://doi.org/10.1016/0022-3697(57)90059-8}
  {\bibfield  {journal} {\bibinfo  {journal} {Journal of Physics and Chemistry
  of Solids}\ }\textbf {\bibinfo {volume} {3}},\ \bibinfo {pages} {121 }
  (\bibinfo {year} {1957})}\BibitemShut {NoStop}%
\bibitem [{\citenamefont {Wei}\ and\ \citenamefont {Chou}(1992)}]{Wei1992}%
  \BibitemOpen
  \bibfield  {author} {\bibinfo {author} {\bibfnamefont {S.}~\bibnamefont
  {Wei}}\ and\ \bibinfo {author} {\bibfnamefont {M.~Y.}\ \bibnamefont {Chou}},\
  }\href {\doibase 10.1103/PhysRevLett.69.2799} {\bibfield  {journal} {\bibinfo
   {journal} {Phys. Rev. Lett.}\ }\textbf {\bibinfo {volume} {69}},\ \bibinfo
  {pages} {2799} (\bibinfo {year} {1992})}\BibitemShut {NoStop}%
\bibitem [{\citenamefont {Huang}, \citenamefont {Voter},\ and\ \citenamefont
  {Perez}(2013)}]{Huang2013}%
  \BibitemOpen
  \bibfield  {author} {\bibinfo {author} {\bibfnamefont {C.}~\bibnamefont
  {Huang}}, \bibinfo {author} {\bibfnamefont {A.~F.}\ \bibnamefont {Voter}}, \
  and\ \bibinfo {author} {\bibfnamefont {D.}~\bibnamefont {Perez}},\ }\href
  {\doibase 10.1103/PhysRevB.87.214106} {\bibfield  {journal} {\bibinfo
  {journal} {Phys. Rev. B}\ }\textbf {\bibinfo {volume} {87}},\ \bibinfo
  {pages} {214106} (\bibinfo {year} {2013})}\BibitemShut {NoStop}%
\bibitem [{\citenamefont {Binder}\ \emph {et~al.}(2015)\citenamefont {Binder},
  \citenamefont {Luskin}, \citenamefont {Perez},\ and\ \citenamefont
  {Voter}}]{Binder2015}%
  \BibitemOpen
  \bibfield  {author} {\bibinfo {author} {\bibfnamefont {A.}~\bibnamefont
  {Binder}}, \bibinfo {author} {\bibfnamefont {M.}~\bibnamefont {Luskin}},
  \bibinfo {author} {\bibfnamefont {D.}~\bibnamefont {Perez}}, \ and\ \bibinfo
  {author} {\bibfnamefont {A.}~\bibnamefont {Voter}},\ }\href {\doibase
  10.1137/140983963} {\bibfield  {journal} {\bibinfo  {journal} {Multiscale
  Modeling \& Simulation}\ }\textbf {\bibinfo {volume} {13}},\ \bibinfo {pages}
  {890} (\bibinfo {year} {2015})},\ \Eprint
  {http://arxiv.org/abs/https://doi.org/10.1137/140983963}
  {https://doi.org/10.1137/140983963} \BibitemShut {NoStop}%
\bibitem [{\citenamefont {Henkelman}\ and\ \citenamefont
  {Jónsson}(2000)}]{neb1}%
  \BibitemOpen
  \bibfield  {author} {\bibinfo {author} {\bibfnamefont {G.}~\bibnamefont
  {Henkelman}}\ and\ \bibinfo {author} {\bibfnamefont {H.}~\bibnamefont
  {Jónsson}},\ }\href {\doibase 10.1063/1.1323224} {\bibfield  {journal}
  {\bibinfo  {journal} {The Journal of Chemical Physics}\ }\textbf {\bibinfo
  {volume} {113}},\ \bibinfo {pages} {9978} (\bibinfo {year} {2000})},\ \Eprint
  {http://arxiv.org/abs/https://doi.org/10.1063/1.1323224}
  {https://doi.org/10.1063/1.1323224} \BibitemShut {NoStop}%
\bibitem [{\citenamefont {Henkelman}, \citenamefont {Uberuaga},\ and\
  \citenamefont {Jónsson}(2000)}]{neb2}%
  \BibitemOpen
  \bibfield  {author} {\bibinfo {author} {\bibfnamefont {G.}~\bibnamefont
  {Henkelman}}, \bibinfo {author} {\bibfnamefont {B.~P.}\ \bibnamefont
  {Uberuaga}}, \ and\ \bibinfo {author} {\bibfnamefont {H.}~\bibnamefont
  {Jónsson}},\ }\href {\doibase 10.1063/1.1329672} {\bibfield  {journal}
  {\bibinfo  {journal} {The Journal of Chemical Physics}\ }\textbf {\bibinfo
  {volume} {113}},\ \bibinfo {pages} {9901} (\bibinfo {year} {2000})},\ \Eprint
  {http://arxiv.org/abs/https://doi.org/10.1063/1.1329672}
  {https://doi.org/10.1063/1.1329672} \BibitemShut {NoStop}%
\bibitem [{\citenamefont {Sheppard}\ \emph {et~al.}(2012)\citenamefont
  {Sheppard}, \citenamefont {Xiao}, \citenamefont {Chemelewski}, \citenamefont
  {Johnson},\ and\ \citenamefont {Henkelman}}]{neb3}%
  \BibitemOpen
  \bibfield  {author} {\bibinfo {author} {\bibfnamefont {D.}~\bibnamefont
  {Sheppard}}, \bibinfo {author} {\bibfnamefont {P.}~\bibnamefont {Xiao}},
  \bibinfo {author} {\bibfnamefont {W.}~\bibnamefont {Chemelewski}}, \bibinfo
  {author} {\bibfnamefont {D.~D.}\ \bibnamefont {Johnson}}, \ and\ \bibinfo
  {author} {\bibfnamefont {G.}~\bibnamefont {Henkelman}},\ }\href {\doibase
  10.1063/1.3684549} {\bibfield  {journal} {\bibinfo  {journal} {The Journal of
  Chemical Physics}\ }\textbf {\bibinfo {volume} {136}},\ \bibinfo {pages}
  {074103} (\bibinfo {year} {2012})},\ \Eprint
  {http://arxiv.org/abs/https://doi.org/10.1063/1.3684549}
  {https://doi.org/10.1063/1.3684549} \BibitemShut {NoStop}%
\bibitem [{\citenamefont {Henkelman}\ and\ \citenamefont
  {Jónsson}(1999)}]{dimer}%
  \BibitemOpen
  \bibfield  {author} {\bibinfo {author} {\bibfnamefont {G.}~\bibnamefont
  {Henkelman}}\ and\ \bibinfo {author} {\bibfnamefont {H.}~\bibnamefont
  {Jónsson}},\ }\href {\doibase 10.1063/1.480097} {\bibfield  {journal}
  {\bibinfo  {journal} {The Journal of Chemical Physics}\ }\textbf {\bibinfo
  {volume} {111}},\ \bibinfo {pages} {7010} (\bibinfo {year} {1999})},\ \Eprint
  {http://arxiv.org/abs/https://doi.org/10.1063/1.480097}
  {https://doi.org/10.1063/1.480097} \BibitemShut {NoStop}%
\bibitem [{\citenamefont {Kadkhodaei}(2018)}]{hesscode}%
  \BibitemOpen
  \bibfield  {author} {\bibinfo {author} {\bibfnamefont {S.}~\bibnamefont
  {Kadkhodaei}},\ }\href@noop {} {\enquote {\bibinfo {title}
  {{lammps-local\_hessian package: A package for LAMMPS software}},}\ } (\bibinfo {year} {2018}),\ \bibinfo
  {note} {\url{https://go.uic.edu/lammps_local_hessian}}\BibitemShut {NoStop}%
\bibitem [{lam(2018)}]{lammps-rebuild}%
  \BibitemOpen
  \href@noop {} {\enquote {\bibinfo {title} {{LAMMPS documentation: Include packages in build}},}\ } (\bibinfo
  {year} {8 Feb 2019}),\ \bibinfo {note}
  {\url{https://lammps.sandia.gov/doc/Build_package.html}}\BibitemShut
  {NoStop}%
\bibitem [{\citenamefont {Bl\"ochl}(1994)}]{paw}%
  \BibitemOpen
  \bibfield  {author} {\bibinfo {author} {\bibfnamefont {P.~E.}\ \bibnamefont
  {Bl\"ochl}},\ }\href {\doibase 10.1103/PhysRevB.50.17953} {\bibfield
  {journal} {\bibinfo  {journal} {Phys. Rev. B}\ }\textbf {\bibinfo {volume}
  {50}},\ \bibinfo {pages} {17953} (\bibinfo {year} {1994})}\BibitemShut
  {NoStop}%
\bibitem [{\citenamefont {Kresse}\ and\ \citenamefont
  {Joubert}(1999)}]{paw_vasp}%
  \BibitemOpen
  \bibfield  {author} {\bibinfo {author} {\bibfnamefont {G.}~\bibnamefont
  {Kresse}}\ and\ \bibinfo {author} {\bibfnamefont {D.}~\bibnamefont
  {Joubert}},\ }\href {\doibase 10.1103/PhysRevB.59.1758} {\bibfield  {journal}
  {\bibinfo  {journal} {Phys. Rev. B}\ }\textbf {\bibinfo {volume} {59}},\
  \bibinfo {pages} {1758} (\bibinfo {year} {1999})}\BibitemShut {NoStop}%
\bibitem [{\citenamefont {Kresse}\ and\ \citenamefont {Hafner}(1993)}]{vasp1}%
  \BibitemOpen
  \bibfield  {author} {\bibinfo {author} {\bibfnamefont {G.}~\bibnamefont
  {Kresse}}\ and\ \bibinfo {author} {\bibfnamefont {J.}~\bibnamefont
  {Hafner}},\ }\href {\doibase 10.1103/PhysRevB.47.558} {\bibfield  {journal}
  {\bibinfo  {journal} {Phys. Rev. B}\ }\textbf {\bibinfo {volume} {47}},\
  \bibinfo {pages} {558} (\bibinfo {year} {1993})}\BibitemShut {NoStop}%
\bibitem [{\citenamefont {Kresse}\ and\ \citenamefont {Hafner}(1994)}]{vasp2}%
  \BibitemOpen
  \bibfield  {author} {\bibinfo {author} {\bibfnamefont {G.}~\bibnamefont
  {Kresse}}\ and\ \bibinfo {author} {\bibfnamefont {J.}~\bibnamefont
  {Hafner}},\ }\href {\doibase 10.1103/PhysRevB.49.14251} {\bibfield  {journal}
  {\bibinfo  {journal} {Phys. Rev. B}\ }\textbf {\bibinfo {volume} {49}},\
  \bibinfo {pages} {14251} (\bibinfo {year} {1994})}\BibitemShut {NoStop}%
\bibitem [{\citenamefont {Kresse}\ and\ \citenamefont
  {Furthmüller}(1996)}]{vasp3}%
  \BibitemOpen
  \bibfield  {author} {\bibinfo {author} {\bibfnamefont {G.}~\bibnamefont
  {Kresse}}\ and\ \bibinfo {author} {\bibfnamefont {J.}~\bibnamefont
  {Furthmüller}},\ }\href {\doibase
  https://doi.org/10.1016/0927-0256(96)00008-0} {\bibfield  {journal} {\bibinfo
   {journal} {Computational Materials Science}\ }\textbf {\bibinfo {volume}
  {6}},\ \bibinfo {pages} {15 } (\bibinfo {year} {1996})}\BibitemShut {NoStop}%
\bibitem [{\citenamefont {Kresse}\ and\ \citenamefont
  {Furthm\"uller}(1996)}]{vasp4}%
  \BibitemOpen
  \bibfield  {author} {\bibinfo {author} {\bibfnamefont {G.}~\bibnamefont
  {Kresse}}\ and\ \bibinfo {author} {\bibfnamefont {J.}~\bibnamefont
  {Furthm\"uller}},\ }\href {\doibase 10.1103/PhysRevB.54.11169} {\bibfield
  {journal} {\bibinfo  {journal} {Phys. Rev. B}\ }\textbf {\bibinfo {volume}
  {54}},\ \bibinfo {pages} {11169} (\bibinfo {year} {1996})}\BibitemShut
  {NoStop}%
\bibitem [{\citenamefont {Perdew}(1992)}]{PERDEW1992}%
  \BibitemOpen
  \bibfield  {author} {\bibinfo {author} {\bibfnamefont {J.~P.}\ \bibnamefont
  {Perdew}},\ }\href {\doibase http://dx.doi.org/10.1016/0375-9601(92)91058-Y}
  {\bibfield  {journal} {\bibinfo  {journal} {Physics Letters A}\ }\textbf
  {\bibinfo {volume} {165}},\ \bibinfo {pages} {79 } (\bibinfo {year}
  {1992})}\BibitemShut {NoStop}%
\bibitem [{\citenamefont {Williams}, \citenamefont {Mishin},\ and\
  \citenamefont {Hamilton}(2006)}]{eamPot}%
  \BibitemOpen
  \bibfield  {author} {\bibinfo {author} {\bibfnamefont {P.~L.}\ \bibnamefont
  {Williams}}, \bibinfo {author} {\bibfnamefont {Y.}~\bibnamefont {Mishin}}, \
  and\ \bibinfo {author} {\bibfnamefont {J.~C.}\ \bibnamefont {Hamilton}},\
  }\href {http://stacks.iop.org/0965-0393/14/i=5/a=002} {\bibfield  {journal}
  {\bibinfo  {journal} {Modelling and Simulation in Materials Science and
  Engineering}\ }\textbf {\bibinfo {volume} {14}},\ \bibinfo {pages} {817}
  (\bibinfo {year} {2006})}\BibitemShut {NoStop}%
\bibitem [{\citenamefont {Fan}\ \emph {et~al.}(2013)\citenamefont {Fan},
  \citenamefont {Huang}, \citenamefont {Yang}, \citenamefont {Raju},
  \citenamefont {Datta}, \citenamefont {Shenoy}, \citenamefont {van Duin},
  \citenamefont {Zhang},\ and\ \citenamefont {Zhu}}]{reaxffLiSi}%
  \BibitemOpen
  \bibfield  {author} {\bibinfo {author} {\bibfnamefont {F.}~\bibnamefont
  {Fan}}, \bibinfo {author} {\bibfnamefont {S.}~\bibnamefont {Huang}}, \bibinfo
  {author} {\bibfnamefont {H.}~\bibnamefont {Yang}}, \bibinfo {author}
  {\bibfnamefont {M.}~\bibnamefont {Raju}}, \bibinfo {author} {\bibfnamefont
  {D.}~\bibnamefont {Datta}}, \bibinfo {author} {\bibfnamefont {V.~B.}\
  \bibnamefont {Shenoy}}, \bibinfo {author} {\bibfnamefont {A.~C.~T.}\
  \bibnamefont {van Duin}}, \bibinfo {author} {\bibfnamefont {S.}~\bibnamefont
  {Zhang}}, \ and\ \bibinfo {author} {\bibfnamefont {T.}~\bibnamefont {Zhu}},\
  }\href {http://stacks.iop.org/0965-0393/21/i=7/a=074002} {\bibfield
  {journal} {\bibinfo  {journal} {Modelling and Simulation in Materials Science
  and Engineering}\ }\textbf {\bibinfo {volume} {21}},\ \bibinfo {pages}
  {074002} (\bibinfo {year} {2013})}\BibitemShut {NoStop}%
\bibitem [{\citenamefont {Fantauzzi}\ \emph {et~al.}(2014)\citenamefont
  {Fantauzzi}, \citenamefont {Bandlow}, \citenamefont {Sabo}, \citenamefont
  {Mueller}, \citenamefont {van Duin},\ and\ \citenamefont
  {Jacob}}]{reaxffPtO}%
  \BibitemOpen
  \bibfield  {author} {\bibinfo {author} {\bibfnamefont {D.}~\bibnamefont
  {Fantauzzi}}, \bibinfo {author} {\bibfnamefont {J.}~\bibnamefont {Bandlow}},
  \bibinfo {author} {\bibfnamefont {L.}~\bibnamefont {Sabo}}, \bibinfo {author}
  {\bibfnamefont {J.~E.}\ \bibnamefont {Mueller}}, \bibinfo {author}
  {\bibfnamefont {A.~C.~T.}\ \bibnamefont {van Duin}}, \ and\ \bibinfo {author}
  {\bibfnamefont {T.}~\bibnamefont {Jacob}},\ }\href {\doibase
  10.1039/C4CP03111C} {\bibfield  {journal} {\bibinfo  {journal} {Phys. Chem.
  Chem. Phys.}\ }\textbf {\bibinfo {volume} {16}},\ \bibinfo {pages} {23118}
  (\bibinfo {year} {2014})}\BibitemShut {NoStop}%
\bibitem [{\citenamefont {Carling}\ \emph {et~al.}(2000)\citenamefont
  {Carling}, \citenamefont {Wahnstr\"om}, \citenamefont {Mattsson},
  \citenamefont {Mattsson}, \citenamefont {Sandberg},\ and\ \citenamefont
  {Grimvall}}]{surfaceEffect}%
  \BibitemOpen
  \bibfield  {author} {\bibinfo {author} {\bibfnamefont {K.}~\bibnamefont
  {Carling}}, \bibinfo {author} {\bibfnamefont {G.}~\bibnamefont
  {Wahnstr\"om}}, \bibinfo {author} {\bibfnamefont {T.~R.}\ \bibnamefont
  {Mattsson}}, \bibinfo {author} {\bibfnamefont {A.~E.}\ \bibnamefont
  {Mattsson}}, \bibinfo {author} {\bibfnamefont {N.}~\bibnamefont {Sandberg}},
  \ and\ \bibinfo {author} {\bibfnamefont {G.}~\bibnamefont {Grimvall}},\
  }\href {\doibase 10.1103/PhysRevLett.85.3862} {\bibfield  {journal} {\bibinfo
   {journal} {Phys. Rev. Lett.}\ }\textbf {\bibinfo {volume} {85}},\ \bibinfo
  {pages} {3862} (\bibinfo {year} {2000})}\BibitemShut {NoStop}%
\bibitem [{\citenamefont {Momma}\ and\ \citenamefont {Izumi}(2011)}]{VESTA}%
  \BibitemOpen
  \bibfield  {author} {\bibinfo {author} {\bibfnamefont {K.}~\bibnamefont
  {Momma}}\ and\ \bibinfo {author} {\bibfnamefont {F.}~\bibnamefont {Izumi}},\
  }\href {\doibase 10.1107/S0021889811038970} {\bibfield  {journal} {\bibinfo
  {journal} {Journal of Applied Crystallography}\ }\textbf {\bibinfo {volume}
  {44}},\ \bibinfo {pages} {1272} (\bibinfo {year} {2011})}\BibitemShut
  {NoStop}%
\bibitem [{\citenamefont {Stukowski}(2010)}]{ovito}%
  \BibitemOpen
  \bibfield  {author} {\bibinfo {author} {\bibfnamefont {A.}~\bibnamefont
  {Stukowski}},\ }\href {http://stacks.iop.org/0965-0393/18/i=1/a=015012}
  {\bibfield  {journal} {\bibinfo  {journal} {Modelling and Simulation in
  Materials Science and Engineering}\ }\textbf {\bibinfo {volume} {18}},\
  \bibinfo {pages} {015012} (\bibinfo {year} {2010})}\BibitemShut {NoStop}%
\bibitem [{\citenamefont {Nakano}(2008)}]{neblammps1}%
  \BibitemOpen
  \bibfield  {author} {\bibinfo {author} {\bibfnamefont {A.}~\bibnamefont
  {Nakano}},\ }\href {\doibase https://doi.org/10.1016/j.cpc.2007.09.011}
  {\bibfield  {journal} {\bibinfo  {journal} {Computer Physics Communications}\
  }\textbf {\bibinfo {volume} {178}},\ \bibinfo {pages} {280 } (\bibinfo {year}
  {2008})}\BibitemShut {NoStop}%
\bibitem [{\citenamefont {Maras}\ \emph {et~al.}(2016)\citenamefont {Maras},
  \citenamefont {Trushin}, \citenamefont {Stukowski}, \citenamefont
  {Ala-Nissila},\ and\ \citenamefont {Jónsson}}]{neblammps2}%
  \BibitemOpen
  \bibfield  {author} {\bibinfo {author} {\bibfnamefont {E.}~\bibnamefont
  {Maras}}, \bibinfo {author} {\bibfnamefont {O.}~\bibnamefont {Trushin}},
  \bibinfo {author} {\bibfnamefont {A.}~\bibnamefont {Stukowski}}, \bibinfo
  {author} {\bibfnamefont {T.}~\bibnamefont {Ala-Nissila}}, \ and\ \bibinfo
  {author} {\bibfnamefont {H.}~\bibnamefont {Jónsson}},\ }\href {\doibase
  https://doi.org/10.1016/j.cpc.2016.04.001} {\bibfield  {journal} {\bibinfo
  {journal} {Computer Physics Communications}\ }\textbf {\bibinfo {volume}
  {205}},\ \bibinfo {pages} {13 } (\bibinfo {year} {2016})}\BibitemShut
  {NoStop}%
\bibitem [{\citenamefont {Tritsaris}\ \emph {et~al.}(2012)\citenamefont
  {Tritsaris}, \citenamefont {Zhao}, \citenamefont {Okeke},\ and\ \citenamefont
  {Kaxiras}}]{diffDFTSi}%
  \BibitemOpen
  \bibfield  {author} {\bibinfo {author} {\bibfnamefont {G.~A.}\ \bibnamefont
  {Tritsaris}}, \bibinfo {author} {\bibfnamefont {K.}~\bibnamefont {Zhao}},
  \bibinfo {author} {\bibfnamefont {O.~U.}\ \bibnamefont {Okeke}}, \ and\
  \bibinfo {author} {\bibfnamefont {E.}~\bibnamefont {Kaxiras}},\ }\href
  {\doibase 10.1021/jp307221q} {\bibfield  {journal} {\bibinfo  {journal} {The
  Journal of Physical Chemistry C}\ }\textbf {\bibinfo {volume} {116}},\
  \bibinfo {pages} {22212} (\bibinfo {year} {2012})},\ \Eprint
  {http://arxiv.org/abs/https://doi.org/10.1021/jp307221q}
  {https://doi.org/10.1021/jp307221q} \BibitemShut {NoStop}%
\bibitem [{\citenamefont {Ostadhossein}\ \emph {et~al.}(2015)\citenamefont
  {Ostadhossein}, \citenamefont {Cubuk}, \citenamefont {Tritsaris},
  \citenamefont {Kaxiras}, \citenamefont {Zhang},\ and\ \citenamefont {van
  Duin}}]{diffusionReaxSi}%
  \BibitemOpen
  \bibfield  {author} {\bibinfo {author} {\bibfnamefont {A.}~\bibnamefont
  {Ostadhossein}}, \bibinfo {author} {\bibfnamefont {E.~D.}\ \bibnamefont
  {Cubuk}}, \bibinfo {author} {\bibfnamefont {G.~A.}\ \bibnamefont
  {Tritsaris}}, \bibinfo {author} {\bibfnamefont {E.}~\bibnamefont {Kaxiras}},
  \bibinfo {author} {\bibfnamefont {S.}~\bibnamefont {Zhang}}, \ and\ \bibinfo
  {author} {\bibfnamefont {A.~C.~T.}\ \bibnamefont {van Duin}},\ }\href
  {\doibase 10.1039/C4CP05198J} {\bibfield  {journal} {\bibinfo  {journal}
  {Phys. Chem. Chem. Phys.}\ }\textbf {\bibinfo {volume} {17}},\ \bibinfo
  {pages} {3832} (\bibinfo {year} {2015})}\BibitemShut {NoStop}%
\end{thebibliography}
%


\end{document}